# Modified Electron Beam Induced Deposition of Metal Nanostructure Arrays using a Parallel Electron Beam

Joysurya Basu and C Barry Carter

Department of Chemical, Materials and Biomolecular Engineering, 191 Auditorium Road, University of Connecticut, Storrs-06269, CT, USA

R Divakar

Physical Metallurgy Division, Indira Gandhi Centre for Atomic Research, Kalpakkam-603102, India

Vijay B Shenoy

Department of Physics, Indian Institute of Science, Bangalore, 560 012 India

N Ravishankar

Materials Research Centre, Indian Institute of Science, Bangalore, 560 012 India

**Abstract**

A modified electron beam induced deposition method using a parallel beam of electrons is developed. The method relies on the build-up of surface potential on an insulating surface exposed to an electron beam. Presence of sharp edges on the insulating surface implies presence of large electric fields that lead to site-specific nucleation of metal vapor on those regions. Feature sizes as small as 20 nm can be deposited without the need to use fine probes and thus the limitation of probe size imposed on the resolution is overcome. The use of pure metal vapor also renders the process inherently clean.

Arrays of metal nanoparticles find applications as catalysts for the growth of arrayed nanotubes/nanowires, as plasmonic waveguides and in high-density recording media [1-3]. A variety of top-down and bottom-up nanofabrication approaches have been developed for producing such arrays [4-11]. Electron-beam-induced deposition (EBID) is one of the favored methods for nanofabrication owing to its site-specificity and applicability for a variety of systems [4,12-16]. However, the resolution of EBID is limited by the probe size and the process is inherently associated with contamination [17]. Here, we report a modified EBID method capable of depositing metal nanoparticle arrays on reconstructed ceramic surfaces over a large area using a parallel beam of electrons. The presence of a space charge layer in an insulator leads to the build-up of an electrostatic potential on its surface. Combined with the topography that consists of sharp edges, an electric field singularity exists along the hill (crest) regions of the reconstructed ceramic surface that can be exploited for site-specific nucleation of metal nanostructures [10,18]. Combined with the fact that this electrostatic driving force can be tuned using an electron beam incident on such a surface, we introduce a method for nanopatterning using a parallel beam of electrons. This enables simultaneous deposition of nanostructures over a large area and the size of the deposits is limited only by the available flux and the exposure time and not by the diameter of the probe. Use of pure metal vapor instead of a carbonaceous precursor gas makes it a clean process. We carry out the deposition inside a TEM that enables in-situ observation of the deposition process.

Annealed m-plane (($10\bar{1}0$) orientation) of alumina was used as the substrate for deposition. Annealing results in the formation of a corrugated surface comprising the ($\bar{1}012$) and the ($10\bar{1}\bar{1}$) facets [19,20]. One half of a thinned TEM sample was coated with 5 nm of Au that serves as the source of the metal. The assembly was heated in the microscope leading to the dewetting of the gold film on the surface. At temperatures close to the melting point of Au (1045 °C), the dewet nanoparticles serve as the source of metal vapor that subsequently deposits on the non-coated side. The deposition takes place in the regions that are exposed to the electron beam, thus enabling direct in-situ observation of the process. A schematic of the process is illustrated in **Fig. 1a**.

*In situ* TEM micrographs of the electron-beam-induced deposition of gold on the reconstructed m-plane recorded after different time intervals of deposition are shown in **Fig.1(b-d)**. It was observed that the deposition starts along the periphery of the beam (lower right in the **Fig. 1b**) and proceeds radially inwards. The linearity of the Au nanostructure is clearly observed with the line direction corresponding to the direction of the crests on the reconstructed surface. Continued exposure of the sample to the electron beam results in evaporation of the Au nanostructure and damage of the substrate that appears as black dots on the substrate.

The deposited samples were analyzed in detail using a field-emission TEM (Tecnai F30) operated at 300 kV. It is observed that the deposition of the metal nanoparticles takes places only on the regions exposed to the electron beam. The deposited regions in the low magnification image shown in (**Fig. 2a**) are indicated by arrows. A higher magnification image of a deposited region is shown in **Fig. 2b**. It is seen from this micrograph that the deposits are elongated in shape (~50 nm long and 20-30 nm wide) and lie parallel to each other. High-resolution TEM image and the corresponding FFT from a deposited region are shown in **Fig. 2c** and 2d respectively. The 0002 planes of sapphire and the Au-sapphire interface can be clearly discerned in this image with the presence of Moiré fringes confirming the presence of Au. In the FFT, the fundamental reflections correspond to those from the sapphire substrate while reflections with smaller spacing in the reciprocal space correspond to the spacing in the Moiré fringes arising out the presence of Au on sapphire (Fig. 2d). The HRTEM/EFTEM images of the deposited regions are also consistent with the presence of Au on the substrate (See Supplementary Information). In some regions, the gold has evaporated from the substrate leading to the formation of faceted pits on the surface of sapphire. The formation of the metal on the 'sharp' crest regions implies that there is driving force for the evaporation on prolonged exposure to high temperatures under low pressure conditions. This can be avoided/minimized by a suitable choice of temperature, pressure and time.

The nucleation of gold takes place preferentially on the hill regions of the structure. Although, it is difficult to infer this directly from the TEM images, SEM investigations on 'ex-situ' samples confirms this fact. Continued exposure leads to the formation of lines completely covering the hill regions in the exposed areas. It is interesting to note that adjacent areas that were not exposed to the electron beam are still *clean* and do not contain any metal lines/arrays. The deposition could be initiated in any region of the substrate by irradiating it with the electron beam. We present below an analysis of the deposition process based on heterogeneous nucleation of the metal on the crest regions. For the case of a symmetric wedge whose surface is at a fixed electrostatic potential, the electric field at a distance 'R' near the crest is given by [21]

$$E = \frac{\sqrt{D}}{R^{\beta}}$$

Where D is a constant that depends upon the magnitude of the potential and the boundary conditions and β is related to the enclosed angle of the crest (α) and is given by

$$\beta = \frac{(\pi - \alpha)}{(2\pi - \alpha)}$$

The radial variation in the electric field near a symmetric dielectric wedge can be illustrated as shown in Fig.3a. The total electrostatic energy per unit length of the crest regions is

$$E_{vac} = \frac{\varepsilon_o}{2} \int_{R_o}^{R_\infty} |E^2| a\rho d\rho = \frac{a\varepsilon_o D}{2} \int_{R_o}^{R_\infty} \rho^{(1-2\beta)} d\rho = a\frac{\varepsilon_o D}{4(1-\beta)}(R_\infty^{2(1-\beta)} - R_o^{2(1-\beta)}) = aC(R_\infty^{2(1-\beta)} - R_o^{2(1-\beta)})$$

Where $\varepsilon_o$ is the electric constant (permittivity of free space), $R_o$ is the radius of the crest, $R_\infty$ is the size of the system, 'a' is defined as $2\pi - \alpha$ and 'C' is defined as $\varepsilon_o D/4(1-\beta)$. If the crest is covered with a metal cylinder of radius R, then the energy per unit length is

$$E_{cyl} = aC \ (R_\infty^{2(1-\beta)} - R^{2(1-\beta)})$$

Here, it is assumed that the parameter 'C' is the same for the bare sapphire surface and one covered with the metal. The change in the electrostatic energy per unit length due to the presence of the liquid cylinder is

$$\Delta E = aC(R_o^{2(1-\beta)} - R^{2(1-\beta)})$$

The total free energy on formation of a liquid cylinder on the ridge (per unit length of the ridge) is given as

$$G = aR\gamma_{LV} + 2R\{\gamma_{SL} - \gamma_{SV}\} - aCR^{2(1-\beta)}$$

Where $\gamma_{LV}$ is the liquid metal-vapor interfacial energy, $\gamma_{SL}$ is the liquid metal-alumina interfacial energy and $\gamma_{SV}$ is the surface energy of alumina and R is determined from the condition that $(a/2)R^2 = v_o$, the total volume of the liquid.

The expression for free energy change is analogous to the free energy change attendant upon heterogeneous nucleation of a liquid on a substrate. In the absence of a driving force for the nucleation of the liquid (undersaturation), the reduction in electrostatic energy plays the role of volume free energy change and aids in the nucleation of the liquid. Although, it is difficult to estimate this parameter 'C' quantitatively for the experimental conditions, it is noted that it depends on the magnitude of the electrostatic potential of the surface. Thus, the driving force for nucleation *can be enhanced by increasing the surface potential* by introducing charges in the specimen through electron irradiation. The effect of changing C on the critical radius for nucleation is shown in **Fig. 3b** where the barrier for nucleation is plotted for two different values of C. Increasing the value of C clearly reduces the barrier as well as the critical radius size and enables nucleation to take place on the crests of the structure. An estimate based on a critical nucleus size of less than 1 nm yields a value of the order of $10^5$ for 'C' for nucleation to take place. This translates to a value of the order of $10^9$ V/m for the electric field (See Supplementary Information). It has been shown in earlier studies that irradiation of insulating surfaces could lead to maximum field strengths that are comparable to this value and thus nucleation can be initiated on the irradiated surface [22]. The effect of high-energy electron beams interacting with an insulating surface has been studied extensively [22-26]. For a thin non-conducting specimen, as in a TEM, the emission of secondary electrons and Auger electrons from the surface leads to a situation where the surface becomes positively charged [24]. The possibility to introduce charges locally in an insulator leads to the possibility of locally changing the surface potential. It has been reported that potential of the order of several tens of volts can be generated on substrates without any topography that translate to maximum electric fields of the order of $10^8 - 10^9$ V/m [22]. It has also been pointed out that the radial component of the field is a maximum at the periphery of the beam. In the present case also, the deposition starts along the periphery of the beam and then proceeds inwards clearly proving the role of electrostatics on the deposition process. On a reconstructed ceramic surface, the fields are a maximum close to the crest regions and hence although irradiation is carried out over a large area, nucleation still takes place selectively along the crest regions present within the irradiated area. In particular, since the vapor phase is undersaturated, nucleation takes place only on the crest regions where the electrostatic driving force is built up due to charging. Thus, this represents a modified high-resolution electron-beam-induced deposition method using a parallel beam of electrons. The limitations of resolution associated with conventional EBID techniques are overcome leading the formation of truly nanoscale structures.

We have demonstrated the formation of nanostructures as small as 10 nm using a parallel beam of electrons from a $LaB_6$ filament in the TEM mode. The resolution of the conventional EBID technique is limited by the probe size [17]. With the use of field-emission sources and operation in the STEM mode, features as small as 1.5 nm have been produced [12,13,15,16]. The resolution in the present method is not limited by the probe size and can be altered by altering the length scale of surface reconstruction and the flux of the metal. In conventional EBID, inorganic or organometallic precursors have been used to deposit metal nanostructures by dissociation of the compound. The dissociated products always contaminate the nanostructures and the deposition environment. The use of pure metal vapor ensures that contamination is significantly reduced in the present method.

NR and CBC acknowledge support through the DST-NSF Indo-US collaborative research program.

1   J.B. Hannon, S. Kodambaka, F.M. Ross, and R.M. Tromp,  Nature  **440**, 69 - 71 (2006).

2   S.A. Maier, P.G. Kik, H.A. Atwater, S Meltzer, E. Harel, B.E. Koel, and A.A.G. Requicha,  Nature Materials **2**, 229-232 (2003).

3   S. Sun, C.B. Murray, D. Weller, L Folks, and A. Moser,  Science **287**, 1989-1992 (2000).

4   A.S Eppler, G Rupprechter, L Guczi, and G.A Somorjai,  J. Phys. Chem. B **101** (48), 9973-9977 (1997).

5   H.O. Jacobs, S.A. Campbell, and M.G. Steward,  Adv. Mater. **14** (21), 1553-1557 (2002).

6   T. R. Jensen, G. C. Schatz, and R. P. Van Duyne,  J. Phys. Chem. B. **103** (13), 2394-2401 (1999).

7   S. Matsui, Y. Kojima, and Y. Ochiai,  Appl. Phys. Lett. **53** (10), 868-870 (1988).

8   R.D. Piner, J. Zhu, F. Xu, S. Hong, and C.A. Mirkin,  Science **283**, 661-663 (1999).

9   C.N.R. Rao, G.U. Kulkarni, P. John Thomas, and Peter P. Edwards,  Chem. Soc. Rev. **29**, 27-35 (2000).

10  N. Ravishankar, Vijay B. Shenoy, and C. Barry Carter,  Adv. Mater. **16** (1), 76-80 (2004).

11  Y Xia and G.M Whitesides,  Ann. Rev. Mat. Sci. **28**, 153-184 (1998).

12  S Frabboni, G.C. Gazzadi, L Felisari, and A Spessot,  Appl. Phys. Lett. **88**, 2131161-2131163 (2006).

13  M Shimojo, K Mitsuishi, M Tanaka, M Han, and K Furuya,  J. Micros. **214** (1), 76-79 (2004).

14  M Song, K Mitsuishi, M Tanaka, M Takeguchi, M Shimojo, and K Furuya,  Appl. Phys. A **80**, 1431-1436 (2005).

[15] M Tanaka, M Shimojo, M Han, K Mitsuishi, and K Furuya, Surf. Interface Anal. **37**, 261-264 (2005).

[16] W.F. van Dorp, B. van Someren, C.W. Hagen, P Kruit, and P.A. Crozier, NanoLetters **5** (7), 1303-1307 (2005).

[17] C Silvis-Cividjian, C.W. Hagen, and P Kruit, J. Appl. Phys. **98**, 0849051-080490512 (2005).

[18] N. Ravishankar and C. Barry Carter, Phil. Mag. Lett. **85** (10), 523-532 (2005).

[19] J. R. Heffelfinger, M. W. Bench, and C. B. Carter, Surf. Sci. Lett. **343**, 1161-1166 (1995).

[20] J.R. Heffelfinger and C.B. Carter, Surf. Sci. **39**, 188–200 (1997).

[21] J.D. Jackson, *Classical Electrodynamics*. (John Wiley and Sons, New York, 1999).

[22] J Cazaux, Ultramicroscopy **60**, 411-425 (1995).

[23] C Attard and J.P. Ganachaud, Phys. Stat. Sol. B **199**, 455-465 (1997).

[24] J Cazaux, J. Appl. Phys. **59** (5), 1418-1430 (1986).

[25] J Cazaux and P Lehuede, J. Electron Spectros. Related Phenomena **59**, 49-71 (1992).

[26] A. Melchinger and S Hofmann, J. Appl. Phys. **78** (10), 6224-6232 (1995).

**Figure Captions**

**Figure 1** (a) TEM micrograph of the reconstructed m-plane sapphire. Thickness fringes and the corrugated diffraction contrast show the reconstructed hill and valley structure of the surface. In situ TEM micrographs of Au deposition with a parallel electron beam after (b) 90 (c) 300 and (d) 630 s of exposure. Continued deposition of nanoparticles leads to the formation of arrayed lines of deposits on the surface.

**Figure 2** (a) Low-magnification image of the deposited regions of the sample. Au deposition takes place only in the regions exposed to the electron beam. (b) Higher-magnification image of the deposited regions. Deposits are elongated in shape and lie parallel to each other. (c) High-resolution image and the corresponding FFT of one of the Au nanostructures. The 0002 planes of sapphire and the Au-sapphire interface can be clearly discerned in this image with the presence of Moiré fringes confirming the presence of Au.

**Figure 3** (a) Distribution of electric field close to a symmetric dielectric wedge (b) Energy barrier for heterogeneous nucleation of metal on a symmetric wedge (in J/unit length of the crest) for different values of surface potential. At higher surface potential, the barrier height reduces and the critical size for heterogeneous nucleation reduces making it favorable for nucleation to take place.

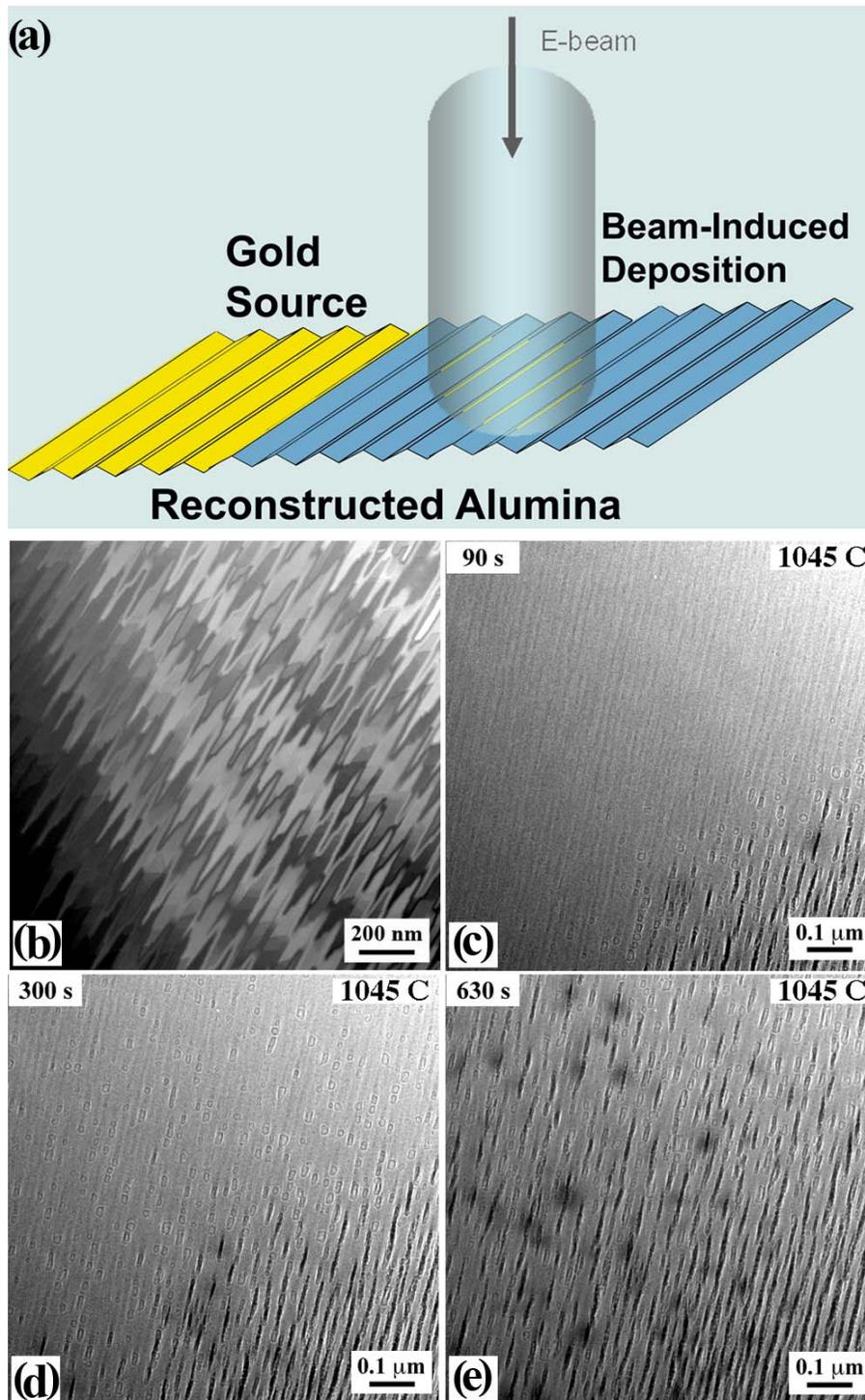

Figure 1 (a) TEM micrograph of the reconstructed m-plane sapphire. Thickness fringes and the corrugated diffraction contrast show the reconstructed hill and valley structure of the surface. In situ TEM micrographs of Au deposition with a parallel electron beam after (b) 90 (c) 300 and (d) 630 s of exposure. Continued deposition of nanoparticles leads to the formation of arrayed lines of deposits on the surface.

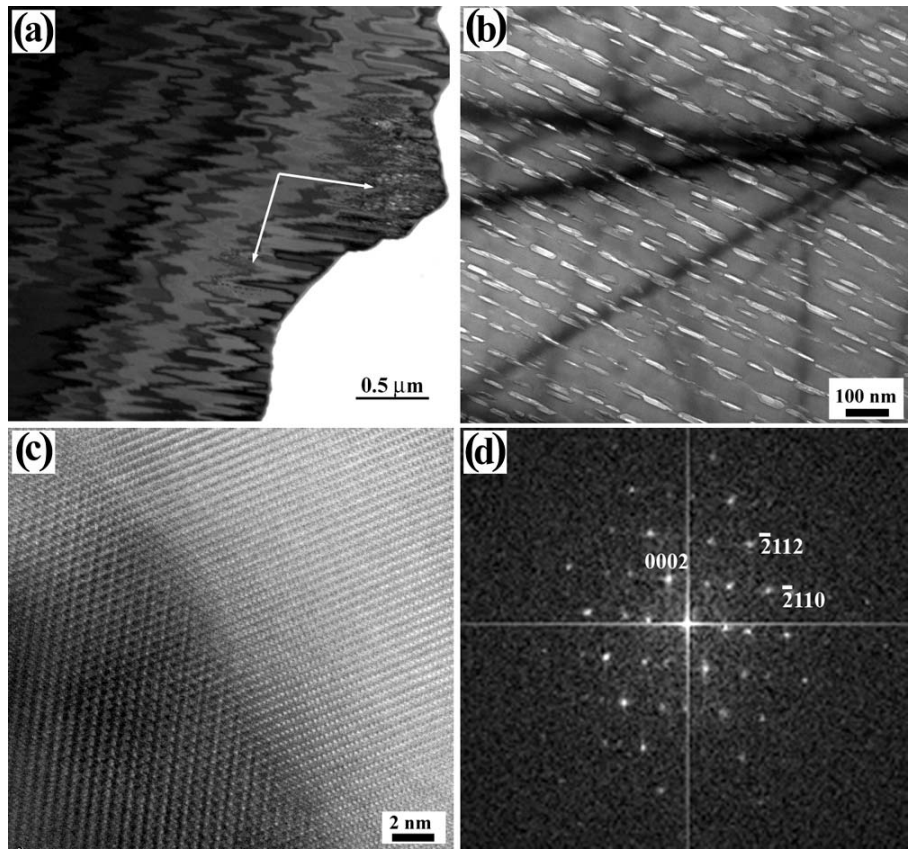

Figure 2 (a) Low-magnification image of the deposited regions of the sample. Au deposition takes place only in the regions exposed to the electron beam. (b) Higher-magnification image of the deposited regions. Deposits are elongated in shape and lie parallel to each other. (c) High-resolution image and the corresponding FFT of one of the Au nanostructures. The 0002 planes of sapphire and the Au-sapphire interface can be clearly discerned in this image with the presence of Moiré fringes confirming the presence of Au.

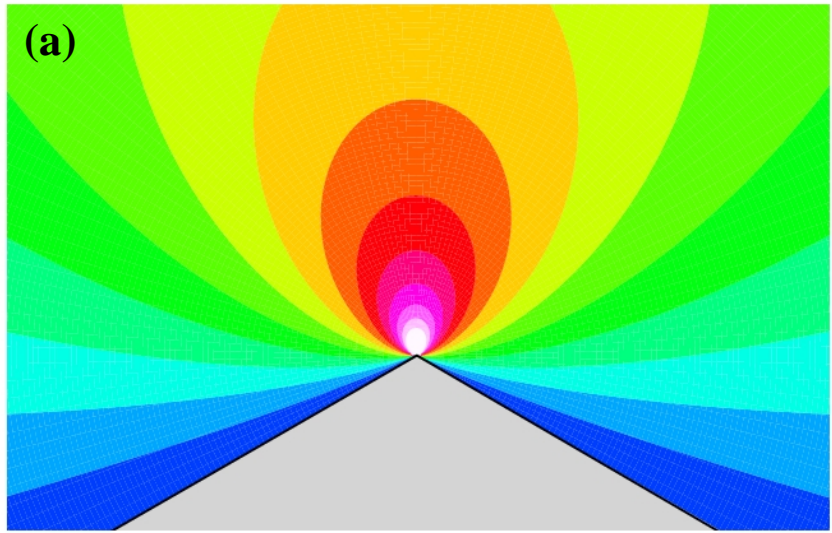

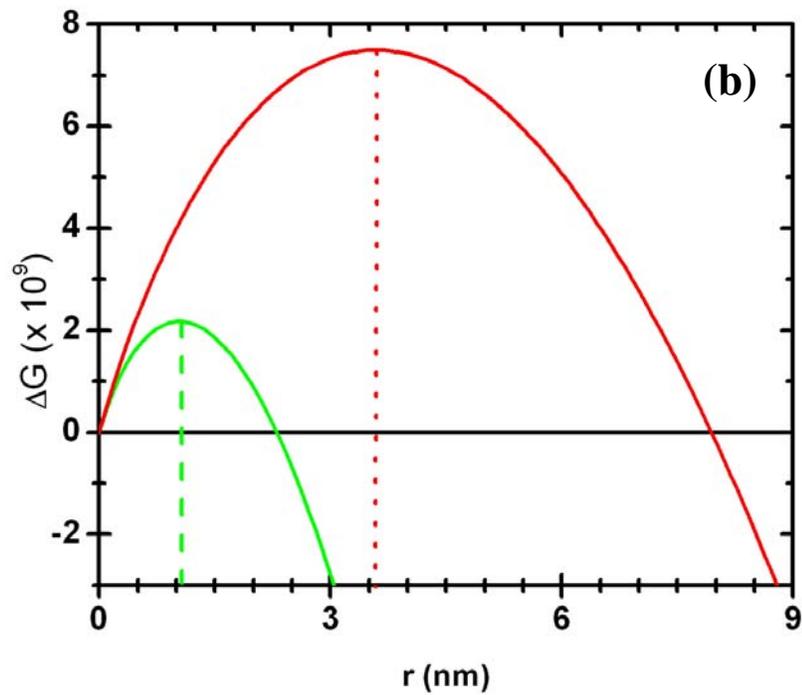

**Figure 3** (a) Distribution of electric field close to a symmetric dielectric wedge (b) Energy barrier for heterogeneous nucleation of metal on a symmetric wedge (in J/unit length of the crest) for different values of surface potential. At higher surface potential, the barrier height reduces and the critical size for heterogeneous nucleation reduces making it favorable for nucleation to take place.